\documentclass[twocolumn,aps,prl,showpacs]{revtex4-1}

\usepackage{times}
\usepackage{graphicx}
\usepackage{amssymb}
\usepackage{amsmath}

\begin{document} 

\title{Contact and sum-rules in a near-uniform Fermi gas at unitarity} 

\author{C. Carcy, S. Hoinka, M. G. Lingham, P. Dyke, C. C. N. Kuhn, H. Hu, and C. J. Vale$^{\ast}$} 

\affiliation{Centre for Quantum and Optical Sciences, ARC Centre of Excellence in Future Low-Energy Electronics Technologies, Swinburne University of Technology, Melbourne 3122, Australia. \\{$^\ast$To whom correspondence should be addressed; E-mail:  cvale@swin.edu.au} }

\begin{abstract}

We present an experimental study of the high-energy excitation spectra of unitary Fermi gases. Using focussed beam Bragg spectroscopy, we locally probe atoms in the central region of a harmonically trapped cloud where the density is nearly uniform, enabling measurements of the dynamic structure factor for a range of temperatures both below and above the superfluid transition. Applying sum-rules to the measured Bragg spectra, we resolve the characteristic behaviour of the universal contact parameter, ${\cal C}$, across the superfluid transition. We also employ a recent theoretical result for the kinetic (second-moment) sum-rule to obtain the internal energy of gases at unitarity.


\end{abstract}


\date{\today}

\maketitle 

Inelastic scattering has revealed a great deal about the structure and properties of matter across diverse areas of physics. High-energy electron-proton scattering yielded evidence for the quark model of nucleons~\cite{Bloom69hei,Breidenbach69obo} and the scattering of fast neutrons from superfluid helium provided crucial input for evaluating the condensate fraction~\cite{Sears82nsd,Griffin93eia}. In cold-atom experiments, inelastic two-photon Bragg spectroscopy has been used to determine the coherence length of Bose-Einstein condensates (BECs)~\cite{Stenger99bso}, map the Bogoliubov spectrum for weakly interacting BECs \cite{Steinhauer02eso} and show behaviours beyond Bogoliubov theory in strongly interacting Bose gases \cite{Papp08bso,Lopes17qei}. In Fermi gases Bragg scattering enabled measurements of the momentum distribution in the non-interacting limit~\cite{Deh09bsa}, as well as, both the density~\cite{Veeravalli08bso} and spin \cite{Hoinka12dsr} response functions of strongly interacting gases near a Feshbach resonance. At high momentum, the dynamic and static structure factors display universal features~\cite{Combescot06msi,Son10sda,Hu10ssf,Hofmann11crs,Nishida12psi,Hofmann17dis} that have been studied in harmonically trapped (inhomogeneous) Fermi gases \cite{Kuhnle10ubo,Hoinka12dsr,Hoinka13pdo}. Central to these universal results is the contact parameter, ${\cal C}$, which defines the probability of finding two interacting atoms with very small separation ~\cite{Tan08eoa}.

Various theoretical studies of the contact in a homogeneous unitary Fermi gas have yielded dramatically different predictions, particularly around the superfluid transition temperature $T_c$ ~\cite{Palestini10tac,Hu11uco,Enss11vas,Drut12ilo,Goulko16nso,Rossi18cam}. Previous attempts to measure this experimentally have either been density-averaged \cite{Partridge05mpo,Werner09noc,Stewart10vou,Kuhnle11tdo,Hoinka13pdo}, or, probed the contact locally \cite{Navon10teo,Sagi12mot,Laurent17cfb}, but to date these have not clearly revealed how the contact evolves across the superfluid transition. Recent progress in the production of homogeneous Bose \cite{Gaunt13bec} and Fermi gases \cite{Mukherjee17haf} provides a promising avenue for studies of critical phenomena~\cite{Navon15cdo} in quantum gases. Density-resolved measurements can also yield local properties, which can be mapped onto an equivalent homogeneous system, provided they satisfy the local density approximation \cite{Ho10otp,Nascimbene10ett,Ku12rts,Lingham14loo}. 

In this letter, we present measurements of the temperature dependence of the contact in a unitary Fermi gas that reveal the characteristic evolution across $T_c$. Our measurements are based on Bragg spectroscopy, using two tightly-focussed laser beams that intersect in the center of a harmonically trapped Fermi gas, where the density is near-uniform \cite{Hoinka17gma}. Bragg spectroscopy yields the dynamic structure factor $S({\mathbf k},\omega)$ where $\hbar {\mathbf k}$ and $\hbar \omega$ are the respective momentum and energy imparted by the two-photon Bragg transition. In an isotropic system $S({\mathbf k},\omega) \equiv S(k,\omega)$, where $k = |{\mathbf k}|$. By employing sum-rules for the dynamic structure factor we determine the static structure factor $S(k)$ and contact \cite{Hu10ssf,Kuhnle10ubo}. Empirically, the high-frequency behaviour of $S(k,\omega)$ is seen to decay approximately as $(\omega - \omega_r)^{-7/2}$, where $\hbar \omega_r = \hbar^2 k^2/(2m) \equiv \epsilon_r$ is the single-particle recoil energy, for modest values of $\omega \: (\gtrsim 2 \omega_r)$. The amplitude of this tail approaches the universal limit based on the contact~\cite{Son10sda}. We also investigate a recent prediction for the kinetic sum-rule to obtain the internal energy of gases at unitarity~\cite{Hofmann17dis}.

Our experiments begin with evaporatively cooled clouds of $^6$Li atoms in a balanced mixture of the lowest two hyperfine states $| F = 1/2, m_F = \pm 1/2 \rangle$ containing between $N/2 = (3 \relbar 10) \times 10^5$ atoms per spin state. Clouds are confined in a hybrid optical-magnetic trap with harmonic trapping frequencies of $(\omega_x, \omega_y, \omega_z)/(2 \pi) = (70, 55, 24.5) \,$Hz \cite{Hoinka17gma}. An external magnetic field is tuned to 832.2\,G where the $s$-wave scattering length diverges ($a \rightarrow \infty$) and elastic collisions reach the unitarity limit \cite{Zurn13pco}. The cloud temperature and size is controlled by varying the end point of the evaporation. Bragg scattering is achieved by focussing two laser beams to a 15 $\mu$m $1/e^2$ radius, which intersect at an angle of $71.6^{\circ}$ ($\omega_r / (2 \pi) = 101.3 \pm 0.3$ kHz) in the center of a trapped cloud, as shown in Fig.~1(a). Our scheme for addressing atoms in the central, near-homogenous portion of the cloud is similar to the approach used in previous studies of Fermi gases to measure the critical velocity \cite{Miller07cvf} and to map the low-lying excitations at small $k$~\cite{Hoinka17gma}. We define the mean density $\bar{n}$ in the Bragg volume as,
\begin{equation}
\bar{n}=\frac{\int n(\mathbf{r})\,\Omega_{Br}^2\,(\mathbf{r})\,\mathrm{d}^3\mathbf{r}}{\int\Omega_{Br}^2(\mathbf{r})\,\mathrm{d}^3\mathbf{r}}, 
\label{nbar}
\end{equation}  
where $\Omega_{Br}({\mathbf r})$ is the spatially-dependent Rabi frequency of the two-photon Bragg transition (which we determine from the intensity profiles of the two focussed Bragg laser beams) and $n(\mathbf{r})$ is the three-dimensional atom density obtained by an inverse Abel transform~\cite{Pretzler91anm}. In the experiments presented here $\bar{n}$ is typically $\sim 0.95 n_0$, where $n_0$ is the peak density of the cloud in the trap center. The mean density sets the relevant momentum and energy scales via the Fermi wavevector, $k_F = (3 \pi^2 \bar{n})^{1/3}$, and Fermi energy $E_F = \hbar^2 k_F^2 / (2m) = k_B T_F$, where $k_B$ is Boltzmann's constant.

\begin{figure}[ht]
\begin{center}
\includegraphics [width=0.47 \textwidth]{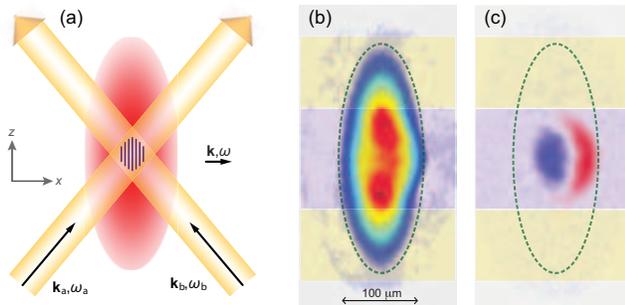} 
\caption{(a) Experimental setup for focussed beam Bragg spectroscopy. Two laser beams with wavevectors $\mathbf{k}_a$ and $\mathbf{k}_b$ and frequencies $\omega_a$ and $\omega_b$, are focussed into the center of a trapped atom cloud. The beams have a $1/e^2$ radius of 15 $\mu$m and intersect at an angle of $2 \theta = 71.6^{\circ}$. (b) Image of a cloud after application of a Bragg pulse for $\omega = (2 \pi) \times$ 50 kHz ($ \approx \omega_r/2$), showing how the central part of the cloud is deformed. (c) Difference between images of clouds with $\omega/(2 \pi)$ = +50 kHz and $\omega/(2 \pi)$ = 0 kHz, used to identify the regions containing scattered atoms for processing. The momentum transferred to the cloud is proportional to the difference between the center of mass of the scattered region (shaded light blue) and the reference region (yellow).}
\end{center}
\label{fig1}       
\end{figure}

The Bragg lasers are detuned approximately $1 \,$THz above the frequency of the nearest atomic transition to probe the density-density response \cite{Hoinka12dsr} while avoiding spontaneous emission. A $100 \, \mu$s Bragg pulse is applied to the trapped cloud, then the trap is turned off and the cloud left to expand for 1 ms before taking an absorption image, Fig.~1(b). From these images, the center of mass of the central part of the cloud $X_{c}$ can be determined, by integrating the light blue shaded region over $z$ and evaluating the first moment. The centers of mass of the two wing regions ($X_{w}$, shaded light yellow) are found in the same way, averaged and subtracted from $X_{c}$ to give the resultant centre of mass displacement, $\Delta X$. To identify the regions to be used for evaluating $X_{c}$ and $X_{w}$ we subtract images obtained using $\omega = 0$ from those obtained at non-zero $\omega$ as shown in Fig.~1(c). Performing the measurement in this way eliminates sensitivity to shot-to-shot fluctuations in the cloud position as both the signal $X_{c}$ and reference $X_{w}$ are obtained from a single image. This sequence is repeated as the Bragg frequency, $\omega = \omega_a - \omega_b$, is varied between $\pm \, 2 \pi \times (0 \relbar 260) \,$kHz to obtain a Bragg spectrum. Both positive and negative Bragg frequencies are used and averaged to improve signal-to-noise.  

We have measured a series of Bragg spectra for clouds at unitarity for temperatures between $0.07 \leq T/T_F \leq 1.1$. Within linear response the rate at which momentum is imparted to the cloud is given by the imaginary part of the dynamic susceptibility 
\begin{equation}
\frac{d\mathbf P}{dt}=-\frac{\hbar \mathbf{k}\,\Omega^2_{Br}}{2}\, {\chi''}(k,\omega), 
\label{BraggProbe}
\end{equation}  
where ${\chi}(k,\omega)$ is the Fourier transform of the retarded density-density correlation function ${\chi}({\mathbf r}-{\mathbf r}',t-t')=-i\theta(t-t')\langle[\hat{n}({\mathbf r},t),\hat{n}({\mathbf r}',t')]\rangle$ \cite{Pines66tto,Brunello01mtt}. This is related to the dynamic structure factor via the detailed balance relation,
\begin{equation}
\chi''(k,\omega) = \pi [ S(k,\omega) - S(-k,-\omega) ].
\label{skw}
\end{equation}  
At high momentum and temperatures satisfying $k_B T \ll \hbar \omega$ only the first term on the right of Eq.~(3) contributes, since high-$k$ states $(\gg k_F)$, relevant for the second term, will be unpopulated. Our measurements thus probe the dynamic structure factor directly.

\begin{figure}[ht]
\begin{center}
\includegraphics [width=0.483\textwidth]{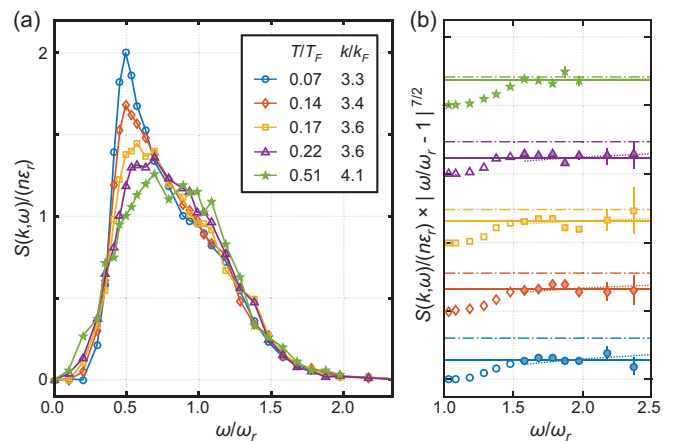} 
\caption{(a) Bragg spectra showing the dynamic structure factor $S(k,\omega)$ for a selection of temperatures above and below the superfluid transition temperature. Relative temperature ($T/T_F$) and Bragg wavevector ($k/k_F$) for each spectrum are shown in the inset. (b) The high-frequency tails of the spectra in (a) multiplied by $| \omega - \omega_r |^{-7/2}$. Solid lines are fits to the tails (filled data points) and dash-dotted lines indicate the predicted tail, Eq.~\ref{tail}, according to the measured contact (displayed in Fig.~3). Dotted lines show a modified fit to the tail, Eq.~(\ref{hitail}), that enforces the expected $\omega \rightarrow \infty$ behavior, as described in the text. }
\end{center}
\label{fig2}       
\end{figure}

A selection of Bragg spectra are shown in Fig.~2(a) for temperatures below and above the superfluid transition, $T_c \approx 0.17 \, T_F$ \cite{Ku12rts}. As with measurements on trapped gases, the coldest spectra are dominated by a peak at half the atomic recoil frequency corresponding to the scattering of pairs from the condensate \cite{Combescot06msi,Lingham14loo}. Above $T_c$ the sharp feature corresponding to pair scattering disappears and the spectral weight shifts to higher frequencies, approaching $\omega_r$. 

In the limit $k \rightarrow \infty$, a universal expression for dynamic structure factor can be found using the operator product expansion (OPE) \cite{Son10sda,Hofmann11crs,Hu12uds,Hofmann17dis}. At high frequencies $S(k,\omega)$ scales with $\omega^{-7/2}$ according to, 
\begin{equation}
S(k,\omega)/(n \epsilon_r) = \frac{16\sqrt{2}}{45 \pi^2} \frac{k_F}{k} \left ( \frac{\omega_r}{\omega} \right )^{7/2} \left ( \frac{\cal C}{n k_F} \right ).
\label{tail}
\end{equation} 

 
Eq.~(\ref{tail}) is also valid at lower $k$ \cite{Taylor10vos,Hofmann11crs} provided the OPE scaling variable $Z = (\omega / \omega_r -1)$ satisfies $Z \gg 1$ \cite{Hofmann17dis}. We find empirically that the high frequency tails of our Bragg spectra are well described by a $Z^{-7/2}$ dependence, both below and above $T_c$, as can be seen in Fig.~2(b). For frequencies $\omega > 1.5 \, \omega_r$, we fit the amplitude of the $Z^{-7/2}$ tail, shown by the solid lines in Fig.~2(b). At the lowest temperatures, the fitted amplitudes, $a_{(ex.)}$, generally lie below the values derived from the measured contact (see below), Eq.~(\ref{tail}), $a_{({\cal C})} \equiv \left (\frac{16\sqrt{2}}{45 \pi^2} \right ) \frac{\cal C}{n k}$, dash-dotted lines in Fig ~2(b). Nonetheless, this approximate $Z^{-7/2}$ dependence suggests universal short-range correlations begin to appear in this energy range.

Energy-weighted moments of the dynamic structure factor, 
\begin{equation}
m_i = \hbar^{i+1} \int_{-\infty}^\infty \omega^i\, S(k,\omega) \, d \omega,
\label{mi}
\end{equation}  
provide additional constraints on the bulk properties of the gas through sum-rules \cite{Pines66tto}. We utilize the zeroth, first and second moments, that define the static structure factor, $f$-sum rule and kinetic sum rule, respectively. For frequencies higher than $2.5 \, \omega_r$ the Bragg response falls below our measurement sensitivity, however, for higher order moments ($i \geq 1$) the tail can carry significant weight. To include this, we assume $S(k,\omega) = a_{(ex.)} / Z^{7/2}$ for $2 \omega_r < \omega < \infty$ in Eq.~(\ref{mi}).

The $f$-sum rule, $m_1 = n \epsilon_r$, valid for all $k$ \cite{Pines66tto,Pitaevskii16bec}, allows a convenient normalisation of the Bragg spectra yielding the dynamic structure factor in units of $n \epsilon_r$, as in Fig.~2(a) \cite{Kuhnle10ubo}. In the large-$k$ limit, the static structure factor can be used to determine the contact \cite{Hu10ssf,Hofmann17dis},
\begin{equation}
\frac{m_0}{m_1} = \frac{S(k)}{\epsilon_r} = \frac{1}{\epsilon_r} \left (1 + \frac{\cal C}{4 n k}  \left [ 1 - \frac{4}{\pi k a} \right ] \right ). 
\label{sk}
\end{equation}  
Using the ratio of the moments, we obtain the dimensionless contact, ${\cal C}/(n k_F) =  4 \frac{k}{k_F} (\epsilon_r \frac{m_0}{m_1} - 1)$, for all of our Bragg spectra, as shown in Fig.~3 (blue circles). Also plotted are various theoretical calculations using a $t$-matrix approach \cite{Palestini10tac}, self-consistent Luttinger-Ward theory \cite{Enss11vas}, Gaussian pair-fluctuation theory (GPF) \cite{Hu11uco}, quantum Monte Carlo (QMC) \cite{Drut12ilo,Goulko16nso} and bold-diagrammatic Monte Carlo (BDMC) \cite{Rossi18cam}. Also shown are previous experimental measurements \cite{Navon10teo,Sagi12mot,Laurent17cfb} of the homogeneous contact. Our data show a clear trend; in the superfluid phase, the dimensionless contact density ${\cal C}/(n k_F)$ starts off near 3 at low $T$ and then drops abruptly to around 2.5 near the critical temperature. Above $T_c$ the contact appears to be relatively stable, decreasing slowly up to $T/ T_F \approx 1$. The error bars on our data are dominated by systematic uncertainties in the determination of the density (based on the inverse Abel transform \cite{Pretzler91anm}). As such, we expect the qualitative shape of this curve to be robust and relatively insensitive to these systematics. Our results are in reasonable agreement with previous measurements \cite{Navon10teo,Sagi12mot, Laurent17cfb} and have a similar shape to the Luttinger-Ward calculation \cite{Enss11vas}. At high temperature, our data approach the virial expansion result (solid dark blue line) \cite{Hu11uco}, albeit with a relatively large error bar. 

\begin{figure}[ht]
\begin{center}
\includegraphics [width=0.483\textwidth]{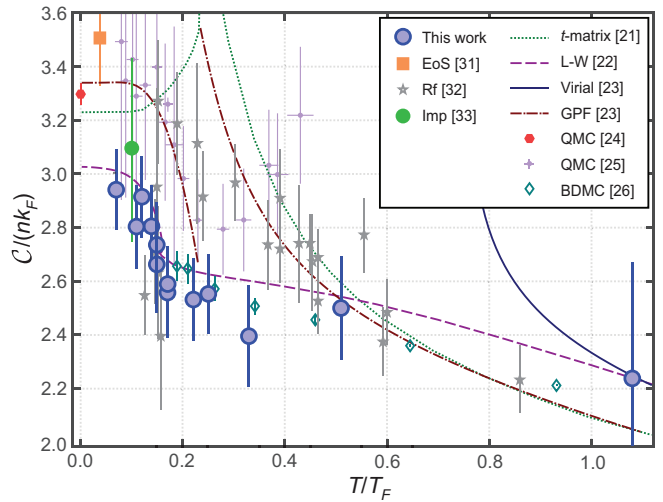} 
\caption{Temperature dependence of the contact parameter ${\cal C}/(n k_F)$ in a Fermi gas at unitarity. Blue filled circles are our experimental data, the orange square is obtained from the pressure equation of state (EoS)~\cite{Navon10teo}, grey stars are previous rf spectroscopy measurements~\cite{Sagi12mot} and the light green circle is obtained from the inelastic loss-rate due to impurity scattering (Imp)~\cite{Laurent17cfb}. Also shown are various theoretical calculations \cite{Palestini10tac,Enss11vas,Hu11uco,Drut12ilo,Goulko16nso,Rossi18cam} (see text for details).}
\end{center}
\label{fig3}       
\end{figure}

At unitarity, a high-$k$ result for the kinetic sum rule was recently derived in terms of the energy density, ${\cal E} \equiv E/V$, where $E$ is the internal energy and $V$ is the volume \cite{Hofmann17dis},
\begin{equation}
\frac{m_2}{m_1} =\epsilon_r + \frac{4}{3} \frac{\cal E}{n}.
\label{kinetic}
\end{equation}  
Rearranging Eq.~(\ref{kinetic}) gives ${\cal E}/{\cal E}_0 = \frac{5}{4} \frac{{\epsilon_r}}{E_F} (\frac{m_2}{m_1} - 1)$, where ${\cal E}_0 = \frac{3}{5} n E_F$ is the energy density of an ideal Fermi gas at zero temperature. In Fig.~4 we plot ${\cal E}/{\cal E}_0$ (blue circles) along with a previous measurement based on the thermodynamic equation of state (dark grey line) \cite{Ku12rts}. The error bars are quite large here, due to the uncertainty in the fitted tail amplitude, $a_{(ex.)}$, which now provides a much larger contribution due to the increased weight of the high frequency points in $m_2$, combined with the uncertainty in the density. 

\begin{table}
\begin{tabular}{ | c | c | c | c | c | }
\hline 
   \; $T/T_F$ \; & \; $k/k_F$ \; & \; $a_{(ex.)}$ \; & \; $a_{({\cal C})}$ \; & \; $b$ \;  \\ \hline 
    0.07(1) & 3.3(1) & 0.024(4) & 0.046 & 1.00 \\
    0.11 & 3.3 & 0.025(4) & 0.044 & 0.90 \\
    0.12 & 3.3 & 0.025(4) & 0.045 & 0.91 \\
    0.14 & 3.4 & 0.025(4) & 0.042 & 0.80 \\
    0.15 & 3.5 & 0.027(4) & 0.041 & 0.49 \\
    0.15 & 3.4 & 0.025(6) & 0.040 & 0.59 \\
    0.17 & 3.5 & 0.022(2) & 0.038 & 0.81 \\
    0.17 & 3.6 & 0.022(5) & 0.037 & 0.69 \\
    0.22(2) & 3.6 & 0.019(4) & 0.036 & 0.93 \\
    0.25 & 3.7 & 0.022(2) & 0.035 & 0.74 \\
    0.33 & 3.7 & 0.025(7) & 0.034 & 0.61 \\
    0.51(5) & 4.1 & 0.027(6) & 0.032 & 0.18 \\
    1.1(1) & 6.4 & 0.02(1) & 0.019 & 0.26 \\
\hline
\end{tabular}
\caption{Experimental parameters for the different Bragg spectra including the cloud temperatures, Bragg wavevector, fitted high-frequency $Z^{-7/2}$ tail amplitude, $a_{(ex.)}$, theoretical tail amplitude, $a_{({\cal C})}$ and the fitted amplitude of next order term $b$ in Eq.~(\ref{hitail}). }
\end{table}


\begin{figure}[ht]
\begin{center}
\includegraphics [width=0.47\textwidth]{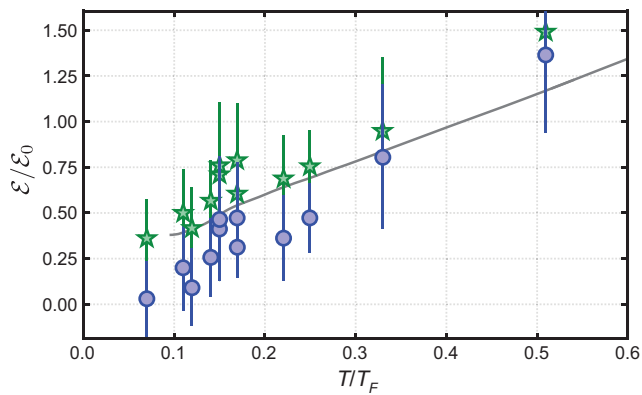} 
\caption{Energy density of a unitary Fermi gas obtained from the $m_2$ sum-rule as function of the temperature. Points are the experimental data based on $m_2/m_1$ where the tail amplitude is set by $a_{(ex.)}$ (blue circles) and for the case where the tail is extrapolated to the $\omega \rightarrow \infty$ limit based on the measured contact (green stars). Grey solid line is the energy obtained from thermodynamic equation of state~\cite{Ku12rts}. }
\end{center}
\label{fig4}       
\end{figure}

At a qualitative level, the data in Fig.~4 show the expected shape, however, at low temperatures our data points lie systematically below the thermodynamic measurement and suggest an unphysical value for the Bertsch parameter $\xi$, which has been found to be $\xi \approx 0.37$~\cite{Ku12rts}. For hotter clouds, our data show better agreement with previous work. The origin of the discrepancy can be traced directly to the amplitude of the fitted $Z^{-7/2}$ tails of the Bragg spectra, Fig.~2(b). Comparing $a_{(ex.)}$ with $a_{({\cal C})}$ in Tab.~1, we see that $a_{(ex.)}$ is nearly a factor of two below $a_{({\cal C})}$ for the lowest temperatures, but approaches $a_{({\cal C})}$ at higher $T$. 

As an alternate approach, we can fit the tail with a function that forces $S(k,\omega)$ to reach the expected limit as $\omega \rightarrow \infty$, given by Eq.~(\ref{tail}). If we assume the tail follows 
\begin{equation}
S(k,\omega \gtrsim 2 \omega_r)/(n \epsilon_r) = \frac{a_{({\cal C})}}{Z^{7/2}} \left [ 1  + \frac{b}{(\omega/\omega_r)} \right ],
\label{hitail}
\end{equation}  
which includes a higher order dependence on $\omega^{-1}$ (the next contributing order at lower $k$~\cite{Hofmann17dis}) and ensures $S(k,\omega \rightarrow \infty) \rightarrow a_{({\cal C})} (\omega / \omega _r)^{-7/2}$, we can fit the parameter $b$ and re-evaluate the moments in a self-consistent way such that the contact and $a_{({\cal C})}$ are set by the new ratio $m_0/m_1$. While the quality of the fit of Eq.~(\ref{hitail}), dotted lines in Fig.~2(b), is essentially equivalent the $Z^{-7/2}$ fit (the sum of the squared residuals differ by $\sim 5\, \%$), this new form for the tail leads to more realistic values of the energy. Green stars in Fig.~4 show the normalised energy density found using Eq.~(\ref{kinetic}) with this new tail. The fitted $b$ coefficients are in the range 0.2 - 1.0 (Tab.~1) and tend to be larger at lower temperatures. The data now show much better agreement with the thermodynamic measurement \cite{Ku12rts}, highlighting the sensitivity of $m_2$ to the high-frequency tail. In contrast, the contact, being set by $m_0$, has only a weak dependence on the precise form of the tail and changes by less than 2$\, \%$ for the two different fit functions, so we do not re-plot this in Fig.~3.

In summary, we have measured the temperature dependence of the universal contact parameter in a near-homogeneous Fermi gas. Our data show that the contact decreases by approximately $15\%$ in the temperature range $0.1 \lesssim T/T_F \lesssim 0.17$. These results establish qualitative and quantitative benchmarks for theories of many-body Fermi systems at finite temperature. We have also checked a recent result for the kinetic sum-rule and found that it can be used to obtain the internal energy, provided the tail of the dynamic structure factor is known out to the asymptotic limit. 

We note that a related study by the group of M. Zwierlein at MIT has found similar results to those presented here. 

We thank W. Zwerger, J. Drut, O. Goulko, G. Strinati and F. Werner for sharing their data and helpful comments on the manuscript. We also thank S. Stringari, M. Zwierlein, B. Mukherjee, and P. Hannaford for valuable discussions. C.J.V. acknowledges financial support from the Australian Research Council programs DP130101807 and CE170100039.


\begin{thebibliography}{9}

\bibitem{Bloom69hei} E. D. Bloom, D. H. Coward, H. DeStaebler, J. Drees, G. Miller, L. W. Mo, R. E. Taylor, M. Breidenbach, J. I. Friedman, G. C. Hartmann, and H. W. Kendall, Phys. Rev. Lett. {\bf 23}, 930 (1969).
\bibitem{Breidenbach69obo} M. Breidenbach, J. I. Friedman, H. W. Kendall, E. D. Bloom, D. H. Coward, H. DeStaebler, J. Drees, L. W. Mo, and R. E. Taylor, Phys. Rev. Lett. {\bf 23}, 935 (1969).
\bibitem{Sears82nsd} V. F. Sears, E. C. Svensson, P. Martel, and A. D. B. Woods, Phys. Rev. Lett. {\bf 49}, 279 (1982).
\bibitem{Griffin93eia} A. Griffin, {\it Excitations in a Bose condensed liquid}. (Cambridge University Press, Cambridge, 1993).
\bibitem{Stenger99bso} J. Stenger, S. Inouye, A. P. Chikkatur, D. M. Stamper-Kurn, D. E. Pritchard, and W. Ketterle, Phys. Rev. Lett. {\bf 82}, 4569 (1999).
\bibitem{Steinhauer02eso} J. Steinhauer, R. Ozeri, N. Katz, and N. Davidson, Phys. Rev. Lett. {\bf 88}, 120407 (2002).
\bibitem{Papp08bso} S. B. Papp, J. M. Pino, R. J. Wild, S. Ronen, C. E. Wieman, D. S. Jin, and E. A. Cornell, Phys. Rev. Lett. 101, 135301 (2008).
\bibitem{Lopes17qei} R. Lopes, C. Eigen, A. Barker, K. G. H. Viebahn, M. Robert-de-Saint-Vincent, N. Navon, Z. Hadzibabic, and R. P. Smith, Phys. Rev. Lett. {\bf 118}, 210401 (2017). 
\bibitem{Deh09bsa} B. Deh, C. Marzok, S. Slama, C. Zimmermann, and P. W. Courteille, Appl. Phys. B {\bf 97} 387 (2009).
\bibitem{Veeravalli08bso} G. Veeravalli, E. Kuhnle, P. Dyke, and C. J. Vale, Phys. Rev. Lett. {\bf 101}, 250403 (2008).
\bibitem{Hoinka12dsr} S. Hoinka, M. Lingham, M. Delehaye, and C. J. Vale, Phys. Rev. Lett. {\bf 109}, 050403 (2012).
\bibitem{Son10sda} D. T. Son, and E. G. Thompson, Phys. Rev. A {\bf 81}, 063634 (2010).
\bibitem{Hu10ssf} H. Hu, X.-J. Liu, and P. D. Drummond, Europhys. Lett. {\bf 91}, 20005 (2010).
\bibitem{Hofmann11crs} J. Hofmann, Phys. Rev. A {\bf 84}, 043603 (2011).
\bibitem{Nishida12psi} Y. Nishida, Phys. Rev. A {\bf 85}, 053643 (2012).
\bibitem{Hofmann17dis} J. Hofmann, and W. Zwerger, Phys. Rev. X {\bf 7}, 011022 (2017).
\bibitem{Combescot06msi} R. Combescot, S. Giorgini, and S. Stringari, Europhys. Lett. {\bf 75}, 695 (2006).
\bibitem{Kuhnle10ubo} E. D. Kuhnle, H. Hu, X.-J. Liu, P. Dyke, M. Mark, P. D. Drummond, P. Hannaford, and C. J. Vale, Phys. Rev. Lett. {\bf 105}, 070402 (2010).
\bibitem{Hoinka13pdo} S. Hoinka, M. Lingham, K. Fenech, H. Hu, C. J. Vale, J. E. Drut, and S. Gandolfi, Phys. Rev. Lett. {\bf 110}, 055305 (2013).
\bibitem{Tan08eoa} S. Tan, Ann. Phys. (N.Y.) {\bf 323}, 2952, {\it ibid.} 2971, {\it ibid.} 2987 (2008).
\bibitem{Palestini10tac} F. Palestini, A. Perali, P. Pieri, and G. C. Strinati, Phys. Rev. A {\bf 82}, 021605(R) (2010).
\bibitem{Enss11vas} T. Enss, R. Haussmann, and W. Zwerger, Ann. Phys. {\bf 326}, 770 (2011).
\bibitem{Hu11uco} H. Hu, X.-J. Liu, and P. D. Drummond, New J. Phys. {\bf 13}, 035007 (2011).
\bibitem{Drut12ilo} J. E. Drut, Phys. Rev. A {\bf 86}, 013604 (2012).
\bibitem{Goulko16nso} O. Goulko, and M. Wingate, Phys. Rev. A {\bf 93}, 053604 (2016).
\bibitem{Rossi18cam} R. Rossi, T. Ohgoe, E. Kozik, N. Prokof’ev, B. Svistunov, K. Van Houcke, and F. Werner, Phys.Rev. Lett. {\bf 121}, 130406 (2018).
\bibitem{Partridge05mpo} G. B. Partridge, K. E. Strecker, R. I. Kamar, M. W. Jack, and R. G. Hulet, Phys.Rev. Lett. {\bf 95}, 020404 (2005).
\bibitem{Werner09noc} F. Werner, L. Tarruell, and Y. Castin, Eur. Phys. J. B {\bf 68}, 401 (2009).
\bibitem{Stewart10vou} J. T. Stewart, J. P. Gaebler, T. E. Drake, and D. S. Jin, Phys. Rev. Lett. {\bf 104}, 235301 (2010).
\bibitem{Kuhnle11tdo} E. D. Kuhnle, S. Hoinka, P. Dyke, H. Hu, P. Hannaford, and C. J. Vale, Phys. Rev. Lett. {\bf 106}, 170402 (2011).
\bibitem{Navon10teo} N. Navon, S. Nascimbene, F. Chevy, and C. Salomon, Science {\bf 328}, 729 (2010).
\bibitem{Sagi12mot} Y. Sagi, T. E. Drake, R. Paudel, and D. S. Jin, Phys. Rev. Lett. {\bf 109}, 220402 (2012).
\bibitem{Laurent17cfb} S. Laurent, M. Pierce, M. Delehaye, T. Yefsah, F. Chevy, and C. Salomon, Phys. Rev. Lett. {\bf 118}, 103403 (2017).
\bibitem{Gaunt13bec} A. L. Gaunt, T. F. Schmidutz, I. Gotlibovych, R. P. Smith, and Z. Hadzibabic, Phys. Rev. Lett. {\bf 110}, 200406 (2013).
\bibitem{Mukherjee17haf}  B. Mukherjee, Z. Yan, P. B. Patel, Z. Hadzibabic, T. Yefsah, J. Struck, and M. W. Zwierlein, Phys. Rev. Lett. {\bf 118}, 123401 (2017).
\bibitem{Navon15cdo} N. Navon, A. L. Gaunt, R. P. Smith, and Z. Hadzibabic, Science {\bf 347}, 6218 (2015).
\bibitem{Ho10otp} T.-L. Ho, and Q. Zhou, Nature Phys. {\bf 6}, 131 (2010).
\bibitem{Nascimbene10ett} S. Nascimbene, N. Navon, K. J. Jiang, F. Chevy, and C. Salomon, Nature {\bf 463}, 1057 (2010).
\bibitem{Ku12rts} M. J. H. Ku, A. T. Sommer, L. W. Cheuk, and M. W. Zwierlein, Science {\bf 335}, 563 (2012).
\bibitem{Lingham14loo} M. G. Lingham, K. Fenech, S. Hoinka, and C. J. Vale, Phys. Rev. Lett. {\bf 112}, 100404 (2014).
\bibitem{Hoinka17gma} S. Hoinka, P. Dyke, M. G. Lingham, J. J. Kinnunen, G. M. Bruun, and C. J. Vale, Nature Phys. {\bf 13}, 943 (2017).
\bibitem{Zurn13pco} G. Z\"{u}rn, T. Lompe, A. N. Wenz, S. Jochim, P. S. Julienne, and J. M. Hutson, Phys. Rev. Lett. {\bf 110}, 135301 (2013).
\bibitem{Miller07cvf} D. E. Miller, J. K. Chin, C. A. Stan, Y. Liu, W. Setiawan, C. Sanner, and W. Ketterle, Phys. Rev. Lett. {\bf 99}, 070402 (2007). 
\bibitem{Pretzler91anm} G. Pretzler, Z. Naturforsch. A {\bf 46}, 629 (1991).
\bibitem{Pines66tto} D. Pines, and P. Nozi\`{e}res, {\it The theory of quantum liquids, Vol. I}, (Benjamin, New York, 1966).
\bibitem{Brunello01mtt} A. Brunello, F. Dalfovo, L. Pitaevskii, S. Stringari, and F. Zambelli, Phys. Rev. A {\bf 64}, 063614 (2001).
\bibitem{Hu12uds} H. Hu, and X.-J. Liu, Phys. Rev. A {\bf 85}, 023612 (2012).
\bibitem{Taylor10vos} E. Taylor, and M. Randeria, Phys. Rev. A {\bf 81}, 053610 (2010).
\bibitem{Pitaevskii16bec} L. Pitaevskii, and S. Stringari, {\it Bose-Einstein condensation and superfluidity}, Oxford University Press, Oxford (2016).






%

\end{thebibliography}
\end{document}